\documentclass[ journal,letterpaper ]{IEEEtran}

\usepackage{graphicx,cite,epsfig,amssymb,amsmath,multicol,subfigure,mathtools,bm,mathrsfs,setspace}
\usepackage{lettrine}
\usepackage{geometry}

\makeatletter
\renewcommand{\maketag@@@}[1]{\hbox{\m@th\normalsize\normalfont#1}}

\begin{document}
\title{\Huge{An Opportunistic-Bit Scheme with IP Styled Communication}}
\author{Bingli Jiao
\thanks{B. Jiao is with the School of Electronics Engineering and Computer Science, Peking University, Beijing, 100871, China (e-mail: jiaobl@pku.edu.cn). He is also with the Joint Laboratory for Advanced Communications between Peking University and Princeton University.}
}


%
%

\maketitle

\begin{abstract}
This work is motivated by the need for the fundamental increase of spectral efficiency with the transmissions on the Transmission Control Protocol and the Internet Protocol (TCP/IP). To emphasize the work in physical layer, we define a bit-unit (BU) that is conceptually similar to an IP packet that contains sufficient information for its destination node to identify the address and interpret the contents in performing the message communication. Armed with these functions, we divide one BU into two parts, which are defined as opportunistic bit (OB) and conventional bit (CB), respectively. In addition, we design the sequential time-slots (TSs) in such a way that the OB can be mapped to the index of a TS, and the CB can be carried by the corresponding TS. To enable the communication, we pre-store a bit-to-TS mapping table at both of the transmitter and the receiver. As result, we can save time resource and gain spectral efficiency as shown in the theoretical analysis confirmed by the simulations.
\end{abstract}

\IEEEpeerreviewmaketitle

\section{Introduction}
This paper reports a new theoretical method suitable to, but not limited in, the Transmission Control Protocol and the Internet Protocol (TCP/IP) packet transmission for increasing spectral efficiency. In a view point of network, the bottleneck of the traffic is, often, found at the channel which is actually shared by a large number of signal resources and receptions in the message communications at the input and output, respectively, e.g., one channel linking two routers is the case.  The congestion occurs whenever the amount of traffic injected into the channel exceeds its capacity.

In IP transmission, we can find that each packet can be correctly received once it departures from the transmitter since the packet contains sufficient information that allows its destination node to identify the addresses and interpret the contents as well.  In addition, the arrivals of those packets do not necessarily be in a time-sequential manner.  Inspired by these, we propose a bit-mapping method that does not necessarily follow the principle of first input and first output in terms of packets at the transmitter. In fact, the order of the transmitted packets is randomized depending on their bit-values.  Moreover, unlike the spatial modulation \cite{SM1,SM2,SM3,SM4} and OFDM index modulation \cite{IM1,IM2,IM3,IM4,IM5}, we do not work neither in the space- nor in subcarrier-index domain, but map some bits in the sequential time-slots (TSs), where the TSs' sequence is fixed while the allocation of the bits is schemed according to the proposed manner.  This yields the concept of the opportunistic bits (OBs) in the communication.

To be broader in sense of the signal methodology, we use the item of ``bit-unit" (termed BU) to replace that of IP ``packet" and build up the model by taking the two advantages: (1) the BU contains sufficient information for its destination to identify and (2) the individual delays do not affect the accomplishments of the message communications. We do not be involved in MAC layer protocols that apply to multi-hop transmission \cite{c1}. In the proposed scheme, every BU is divided into two parts: one is defined as the OB and the other the conventional bit (CB). The former will be mapped to the index of a TS, while the latter is carried by the corresponding TS to each channel realization.  Actually, the successful communication of a BU will attribute to that of each OB and CB as a bit-pair transmitted from the transmitter to the receiver.

The proposed scheme requires the transmitter to accumulate a large number of BUs before triggering its signal transmission. Thus, the traffic is not considered to be interactive or real time.  Meanwhile, this work is restricted in the study on the theoretical limit and, thus, not involved much in the discussions over hardware complexity and costs. Finally, we work in the physical layer over additive Gaussian noise (AWGN) channel for showing a fundamental performance of this approach.

\section{System Model And Analysis}
Figure 1 shows the proposed system having the transmitter, the receiver and the channel that is assumed as AWGN channel.  The scenario is, often, found in TCP/IP transmission for the congestion problem prompting the study of end-to-end control algorithms \cite{TCPIP}. However, we work mostly in physical layer, where communication performance is characterized by spectral efficiency and bit error ratio (BER) performance.

In contrast to the conventional transmitter, as illustrated in Fig. 1, we have three new components which are respectively termed the segmentator, the TS-selector and the storage. Specifically, these components correspondingly fulfill dividing every BU into two parts, performing TS indices mapping, and facilitating the sequential TSs' injection into the channel encoder.

In the system operation, we assume that the number of the BUs is infinitive large and the BUs are independent of each other. Like IP packet transmission, we assume that each BU contains sufficient information for the receiver to identify the address and interpret the messages. Thus, the sequence of the BUs places no role in sense of message communications. In addition, we assume that one channel code can be used to a single BU or shared by a number of BUs jointly.

\begin{figure}[!t]   
\centering
\includegraphics[width=0.45\textwidth]{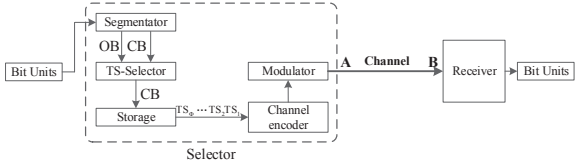}
\caption{The system diagram}
\label{fig1}
\end{figure}

\subsection{Bit-representation}
Assume, temporarily, that all BUs are binary bits of equal length as the inputs of the transmitter.  We write a BU in a vector form as
\begin{eqnarray}
\begin{array}{*{20}{l}}
{{{\bf{g}}^{\left( m \right)}} = \left\{ {g_1^{\left( m \right)},g_2^{\left( m \right)}, \cdots ,g_K^{\left( m \right)},g_{K + 1}^{\left( m \right)}, \cdots ,g_N^{\left( m \right)}} \right\}}\\
{for\;\;m = 1,~2,~\cdots,~M,~M+1, \cdots,}
\end{array}
\end{eqnarray}
where  $m$ denotes the sequential index of the BUs, 
$N$ is the total bit number of one BU, and ${g_i^{\left( m \right)}}$ indicates the $i$-th entry of ${{\bf{g}}^{\left( m \right)}}$, being with two possible values, 0 or 1, of equal probability, $K$ is the factor that labels the point, by which ${{{\bf{g}}^{\left( m \right)}}}$ is divided into two parts, which are defined as the OB and the CB as expressed by the sub-vectors
\begin{subequations}\label{equ2}
\begin{eqnarray}
\begin{array}{l}\label{equ2a}
{\bf{g}}_{OB}^{\left( m \right)} = \left\{ {g_1^{\left( m \right)},g_2^{\left( m \right)}, \cdots ,g_K^{\left( m \right)}} \right\}
\end{array}
\end{eqnarray}
and
\begin{eqnarray}
\begin{array}{l}\label{equ2b}
{\bf{g}}_{CB}^{\left( m \right)} = \left\{ {g_{K + 1}^{\left( m \right)}, \cdots ,g_N^{\left( m \right)}} \right\}
\end{array},
\end{eqnarray}
\end{subequations}respectively, where the OB contains $K$ bits and the CB $\left(N-K\right)$ bits. It is noted that OB and CB jointly hold all information bits of one BU, e.g., ${\bf{g}}_{OB}^{\left( m \right)}$ and ${\bf{g}}_{CB}^{\left( m \right)}$ jointly hold the bits of ${\bf{g}}^{\left( m \right)}$.

For creating the bit-representation of the OB, we prepare the bit-mapping pool by constructing a complete bit-set $\bf V$ with
\begin{eqnarray}
\begin{array}{*{20}{l}}
{{{\bf{v}}^{\left( i \right)}} = \left\{ {v_1^{\left( i \right)},v_2^{\left( i \right)}, \cdots ,v_K^{\left( i \right)}} \right\}\;\;\;\;for\;\;i = 1,2, \cdots ,\Phi }
\end{array},
\end{eqnarray}
where ${{\bf{v}}^{\left( i \right)}} $ is the $i$-th base-vector of $\bf V$ and $\Phi$ is the total number of the base-vectors, satisfying $\Phi  = {2^K}$, and $v_k^{\left( i \right)}$ is the $k$-th entry of ${{\bf{v}}^{\left( i \right)}}$, with the bit-value being either 0 or 1. 
It is found that ${\bf{g}}_{OB}^{\left( m \right)} \in {\bf{V}}$ holds.

We make a mapping table (see Table I) that uniquely connects each base-vector in $\bf V$ to the time sequential index.  For a given ${\bf g}_{OB}^{\left( m \right)}$, we can find its identical base-vector and, then, use Table \ref{Table1} to recode the corresponding TS's index. As a result, we can map each ${\bf g}_{OB}^{\left( m \right)}$ to the recoded TS' index.

To complete the representation of one BU, we load ${\bf g}_{CB}^{\left( m \right)}$ to the corresponding TS, whose index has been determined by the OB, i.e., ${\bf g}_{OB}^{\left( m \right)}$ mentioned above. Finally, we can find that the index of the TS and the contents inside the TS contain all the information bits of the BU.

Moreover, we pre-store Table \ref{Table1} in both the transmitter and the receiver so that the communication can effectively work as explained next.

\newcommand{\tabincell}[2]{\begin{tabular}{@{}#1@{}}#2\end{tabular}}
\begin{table}[htb]
\renewcommand{\arraystretch}{1.5}
\centering
\small
\caption{A Mapping Table of the Complete Bit-Set $\bf V$}
\label{Table1}
\begin{tabular}{|c|c|}
\hline
\tabincell{c}{\textbf {Base-Vector}} & \tabincell{c}{\textbf {TS Sequential Index}} \\
\hline
\tabincell{c}{${{{\bf{v}}^{\left( 1 \right)}}}$} & \tabincell{c}{1} \\
\hline
\tabincell{c}{${{{\bf{v}}^{\left( 2 \right)}}}$} & \tabincell{c}{2} \\
\hline
\tabincell{c}{$.$} & \tabincell{c}{$.$} \\
\hline
\tabincell{c}{$.$} & \tabincell{c}{$.$} \\
\hline
\tabincell{c}{${{{\bf{v}}^{\left( \Phi \right)}}}$} & \tabincell{c}{$\Phi $} \\
\hline
\end{tabular}
\end{table}

\subsection{Communication Scheme}
As shown in Fig. \ref{fig1}, when one BU is taken to the transmitter as the input, the segmentator divides it into the OB and the CB, e.g., ${\bf g}^{\left( m \right)}$ is divided into ${\bf g}_{OB}^{\left( m \right)}$ and ${\bf g}_{CB}^{\left( m \right)}$, having $K$ bits and $N-K$ bits, respectively. The TS-selector takes the OB, i.e., ${\bf g}_{OB}^{\left( m \right)}$, to compare with all base-vectors in $\bf V$ and find the identical base-vector. By utilizing Table \ref{Table1} to find the index of a TS, we load the bit-value of CB, i.e., ${\bf g}_{CB}^{\left( m \right)}$, to the corresponding TS.

\begin{figure}[!t]   
\centering
\includegraphics[width=0.35\textwidth]{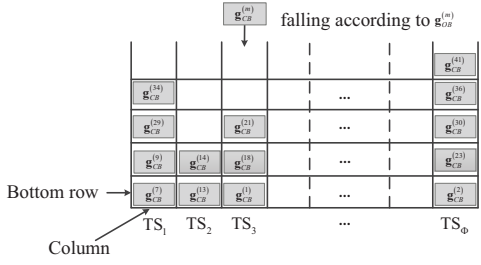}
\caption{The diagram of the falling model}
\label{fig2}
\end{figure}

We create a falling model to work with the OB, the CB and the sequential TSs as shown in Fig. \ref{fig2}, which contains $\Phi$ columns, whose indices are the same as that of the TSs in Table I. We mark TSs as ${{\rm{TS}}_1}, {{\rm{TS}}_2}, \cdots ,{{\rm{TS}}_{\Phi}}$ at the bottom of the storage and find each column contains the layered TS entities, which are used to load CBs. The rows of the storage are used to distinguish the layered TSs in the same column. In the initial state, all TSs have been pre-stored by zero bits.

For a given BU, once the sequential index of a TS has been selected according to ${\bf g}_{OB}^{\left( m \right)}$ by the TS-selector, ${\bf g}_{CB}^{\left( m \right)}$ will be dropped to the corresponding column from the above to the bottom of the storage, being loaded to the corresponding TS. More specifically, loading the bits of the CB is actually to replace all zero bit-values which are pre-stored in the TSs. The storage will be loaded by CB one another. Whenever more than one CB has been dropped in the same column, they will be layered from bottom to above. While, when anyone of the TSs was unloaded by CB, it has been pre-stored by all zero bits whose number is same as that of a CB.

We take the contents in Fig. 2 as an example to explain the above procedure. As can be found, when starting with ${\bf g}^{\left(1\right)}$, the TS-selector maps ${\bf g}_{OB}^{\left(1\right)}$ to a certain TS index, for example to ${\rm {TS}}_3$. Hence the ${\bf g}_{CB}^{\left(1\right)}$ will be dropped to the third column. When continuing with ${\bf g}^{\left(2\right)}$, ${\bf g}_{CB}^{\left(2\right)}$ could be dropped to a different column if ${\bf{g}}_{OB}^{\left( 2 \right)} \ne {\bf{g}}_{OB}^{\left( 1 \right)}$ , but ${\bf g}_{CB}^{\left(2\right)}$ can be dropped to the same column if ${\bf{g}}_{OB}^{\left( 2 \right)} = {\bf{g}}_{OB}^{\left( 1 \right)}$. In the latter case, ${\bf g}_{CB}^{\left(1\right)}$ and ${\bf g}_{CB}^{\left(2\right)}$ will be in the layered shape. We draw the case for ${\bf{g}}_{OB}^{\left( 2 \right)} \ne {\bf{g}}_{OB}^{\left( 1 \right)}$ in Fig. \ref{fig2}. Actually, different OBs will distribute the CBs in different columns, while the same OBs will restrict their CBs in the same column. 

The communication begins when the storage has accumulated $M$ CBs inside.  The first-round injection takes place at the bottom, where $\Phi$ TSs will be injected into the channel encoder in a sequential manner of ${{\rm{TS}}_1}, {{\rm{TS}}_2}, \cdots ,{{\rm{TS}}_{\Phi}}$. During the injection, whenever one TS has been injected to the channel encoder, the above TSs in the same column will immediately fall one row down to guarantee the availability of the second-round injection. Meanwhile, one BU should be processed to have its CB fallen down to the storage so that the storage always contains $M$ CBs inside.
After completing the first-round injection, the second-round injection will start from the first column till the last column. Such injections will be performed repeatedly one round another.

At the receiver, after the demodulation and decoding, the BU can be recovered by the following process. Knowing the length of one BU and the division manner of OB and CB, the receiver can find the indices of the TSs by accounting the received sequential index $\Psi$. In specific, we can find the index of a TS in the formula as
\begin{eqnarray}
\begin{array}{l}\label{equ4}
i = \Psi  - n\Phi \;\;\;\; for \;\;\;\; \Psi=1,~2,~3,~\cdots
\end{array},
\end{eqnarray}
where  $\Psi$ is the received sequential index, $n = \frac{{\left\lfloor {\Psi  - 1} \right\rfloor }}{\Phi }$ indicates the round number, and $i$ is the index of the TS. By using Table \ref{Table1} and reading the content of each TS, every BU can be fully recovered. In the practical considerations, we note that the proposed scheme hold when OB is segmented from any part of the BU. Though we design the OB in front of CB in this paper. 


\subsection{Analysis}

\begin{figure}[!t]   
\centering
\includegraphics[width=0.4\textwidth]{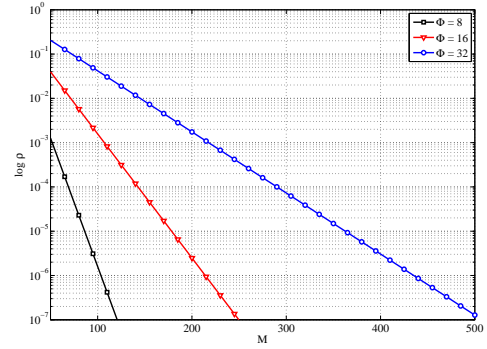}
\caption{The probability versus $M$ for $\Phi  = 8$, $\Phi  = 16$, and $\Phi  = 32$.}
\label{fig3}
\end{figure}

In the above design of the signal transmission, we can note that the BUs are loaded to TSs in a manner of random-falling. Consequently, for loading $M$ BUs, there is a probability that one TS is not loaded by CB, but pre-stored by the zero-bits. We define such an unloaded TS as a TS error at the transmitter. Nevertheless, the error probability can become infinitive small when we increase the number of the loaded BUs, i.e, $M$. Working on a snapshot of Fig. \ref{fig2}, the probability of a CB falling outside of a certain column in the falling model is $\frac{{\Phi  - 1}}{\Phi }$. Thus, for loading $M$  CBs, the probability of a certain column in the falling model being unloaded can be estimated by
\begin{eqnarray}
\begin{array}{l}\label{equ5}
\rho  = {\left( {\frac{{{\Phi-1}}}{\Phi }} \right)^M}
\end{array}.
\end{eqnarray}
We show the numerical results of (\ref{equ5}) in Fig. \ref{fig3} for $\Phi  = 8$, $\Phi  = 16$, and $\Phi  = 32$ corresponding to 3 bits, 4 bits, and 5 bits designed in the OB, respectively. It is found that larger $\Phi$ requires more BUs in the accumulation, i.e., larger $M$, for keeping the same level of the transmission error. However, we work with an assumption of infinitive small probability of the unloading TS by $\rho \approx 0$ in the following discussions since we are interested in the theoretical limit.  

With the help of OB, we can find that the spectral efficiency has been increased by
\begin{eqnarray}
\begin{array}{l}\label{equ6}
\eta  = \frac{N}{{N - K}}
\end{array},
\end{eqnarray}
where $\eta$ is the gain factor of the spectral efficiency. 
One can find that the conclusion of the spectral efficiency gain is true when considering the fact that the use of the channel coding is independent of the proposed scheme. In other words, the bit error performance owing to channel coding does not give any negative impacts to (\ref{equ6}). Actually, the BER performance can be benefited separately as explained in the next section. 

We have to mention a shortcoming of this scheme: in order to achieve high spectral efficiency gain as described in (6), we require a large number of TSs since $\Phi  = {2^K}$. Consequently, the delay will be increased because the storage has to hold a vast number of BUs, i.e, $M$ as found in Fig. 3, before triggering the transmission. In addition, the storage size can also be drastically increased for holding all CBs.

Finally, the above method can be extended to a case where various BUs are with different lengths. In that case, we can categorize the BUs according to their lengths into groups, in each of which the BUs are of same length in number of bits. Then, we can employ a multi-carry system, in which one carry is used to deal with one kind of length and perform similar operations as described above.

\section{Simulation}
In this section, we will perform two simulations to show the advantages of this approach in terms of the signal-to-noise ratio (SNR) gain. The BUs of 36 bits in the length are taken to the simulations by designing 4 OBs and 32 CBs, and BPSK in signal modulation.

The BUs are generated using PC computer with MATLAB functions and, then, taken to the processing as shown in Fig. \ref{fig1}. By setting $M=256$ to the storage, we do the simulations for the code rate of $R = 1$ and $R = \frac{1}{2}$, respectively. The former is equivalent to the case of non-channel coding, whereat we use a hard-decision method to the signal detection. For the latter, we use LDPC code of 1152 bits (with 802.11 standard) in the length with the sum-product algorithm for decoding. Actually, for each channel code, we have transmitted 32 BUs with $32\times4=128$ bits of the OBs by using the TSs' indices and $32\times32=1024$ bits of the CBs taken into the channel encoder to the channel realizations. However, in the conventional scheme, all bits are transmitted by channel realizations in the simulations.

In comparison between the two schemes, we have found the following two merits of the proposed scheme. The first one is the higher data rate because that the saving of OB in channel realizations leads to the spectral efficiency gain indicated in (\ref{equ6}). For each BU, the proposed scheme transmits 32 bits instead of 36 bits with the conventional scheme, since we have saved a time resource of 4 bits owing to the OB. The second is on the power gain because that OBs do not use any power in the transmission. They save 4 bits' energy from 36 bits. This allows a SNR gain of 36/(36-4).

\begin{figure}[!t]   
\centering
\includegraphics[width=0.35\textwidth]{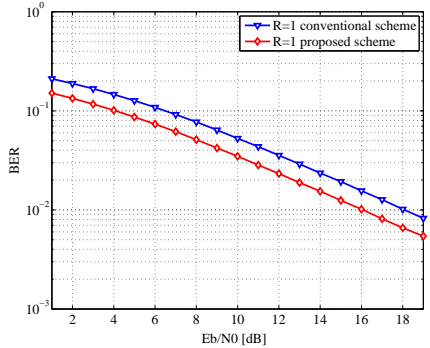}
\caption{The BER comparison between the proposed scheme and the convention method with the code rate of $R=1$.}
\label{fig4}
\end{figure}

\begin{figure}[!t]   
	\centering
	\includegraphics[width=0.35\textwidth]{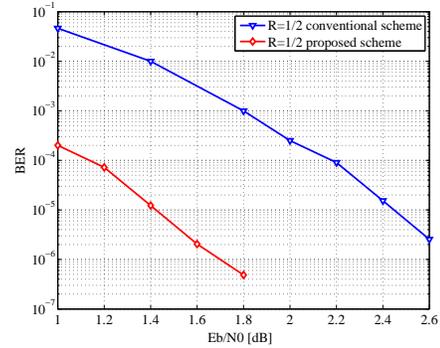}
	\caption{The BER comparison between the proposed scheme and the convention method with the code rate of $R=1/2$.}
	\label{fig5}
\end{figure}

The simulation results are compared in Fig. \ref{fig4} and Fig. \ref{fig5} for the two schemes, i.e., without channel coding and encoded modulation, respectively. In Fig. \ref{fig4}, one can find that the SNR gain exceeds 2 dB in the BER performance in comparison with that of the conventional scheme. In Fig. \ref{fig5}, the SNR gain is higher than 0.96 dB. The phenomenon can be explained by the reason of the SNR gain mentioned above.

\section{Conclusion}
We have proposed a new scheme by creating the OB, at the transmitter, that allows the BU to be shorten in time domain with the channel realizations. To explain the mechanism of the signal transmission, we have proposed a falling model and worked out a theoretical spectral efficiency gain owing to the use of OB for saving the time resource. In addition, we have found the SNR gain in aspect of saving power by the OB. The application of this approach can be incentive due to the two gains. However, the problems of the delay and the storage size should be considered in the future work.

\section{Acknowledgement}
The author thanks his students, Ms.Yingyang Chen and Mr. Shuyi Tian, for their work on the numerical calculations and the figures with the Latex version.


\begin{thebibliography}{1}

%
%
%
%
%
%
%


\bibitem{SM1}
Y. Yang and B. Jiao, ``Information-guided channel-hopping for high data rate wireless communication," \emph{IEEE Communications Letters}, vol. 12, no. 4, pp. 225--227, April 2008.

\bibitem{SM2}
R. Y. Mesleh, H. Haas, S. Sinanovic, C. W. Ahn, and S. Yun, ``Spatial modulation," \emph{IEEE Transactions on Vehicular Technology}, vol. 57, no. 4, pp. 2228--2241, July 2008.

\bibitem{SM3}
D. A. Basnayaka, M. Di Renzo, and H. Haas, ``Massive but few active MIMO," \emph{IEEE Transactions on Vehicular Technology}, vol. 65, no. 9, pp. 6861--6877, Sept. 2016.

\bibitem{SM4}
M. Di Renzo, H. Haas, A. Ghrayeb, S. Sugiura, and L. Hanzo, ``Spatial modulation for generalized MIMO: Challenges, opportunities, and implementation," \emph{Proceedings of the IEEE}, vol. 102, no. 1, pp. 56--103, Jan. 2014.

\bibitem{IM1}
E. Basar, U. Aygolu, E. Panayirci, and H. V. Poor, ``Orthogonal frequency division multiplexing with index modulation," \emph{IEEE Transactions on Signal Processing}, vol. 61, no. 22, pp. 5536--5549, Nov. 2013.

\bibitem{IM2}
T. Datta, H. S. Eshwaraiah, and A. Chockalingam, ``Generalized space-and-frequency index modulation," \emph{IEEE Transactions on Vehicular Technology}, vol. 65, no. 7, pp. 4911--4924, July 2016.

\bibitem{IM3}
T. Mao, Q. Wang, and Z. Wang, ``Generalized dual-mode index modulation aided OFDM," \emph{IEEE Communications Letters}, to be appeared.

\bibitem{IM4}
R. Fan, Y. J. Yu, and Y. L. Guan, ``Generalization of orthogonal frequency division multiplexing with index modulation," \emph{IEEE Transactions on Wireless Communications}, vol. 14, no. 10, pp. 5350--5359, Oct. 2015.

\bibitem{IM5}
T. Mao, Z. Wang, Q. Wang, S. Chen, and L. Hanzo, ``Dual-mode index modulation aided OFDM," \emph{IEEE Access}, vol. 5, pp. 50--60, Feb. 2017.

\bibitem{c1} 
V. Firoiu and M. Borden, ``A study of active queue management for congestion control," in \emph{Proc. 2000 IEEE INFOCOM}, 2000, pp. 1435-1444, vol.3.

\bibitem{TCPIP} 
Uyless Black,  ``TCP/IP and related protocols," Book, McGraw-Hill, 1998.

\end{thebibliography}
\end{document}